\documentclass{article}

\usepackage{arxiv}

\usepackage[T1]{fontenc}
\usepackage[utf8]{inputenc}
\usepackage[english]{babel}
\usepackage{graphicx}
\usepackage{subcaption}
\graphicspath{ {./images/} }
\usepackage{color}
\usepackage{xcolor}
\usepackage{enumerate}
\usepackage{amsfonts, amssymb, amsmath,amsthm,amscd,wasysym}
\usepackage{multirow, array} 
\usepackage{physics}
\usepackage[hidelinks]{hyperref}     
\usepackage{url}                     
\usepackage{booktabs}                
\usepackage{nicefrac}                
\usepackage{microtype}               
\usepackage{lipsum}
\usepackage{authblk}                 

\newcommand{\anti}[1]{\overline{#1}}
\newcommand{\seu}{s_{e\mu}}
\newcommand{\ceu}{c_{e\mu}}
\newcommand{\set}{s_{e\tau}}
\newcommand{\cet}{c_{e\tau}}
\newcommand{\teu}{t_{e\mu}}
\newcommand{\tet}{t_{e\tau}}
\newcommand{\carf}[2]{\left(\frac{#1}{#2}\right)}
\newcommand{\yukawa}[2]{h_{2 #1}^{\nu #2}}

\title{Lepton masses in a non universal U(1) model with three families}

\author[]{
C. Cortes-Parra\thanks{\url{camacortespar@unal.edu.co}}\hspace{3pt},\hspace{5pt}            
R. Martinez\thanks{\url{remartinezm@unal.edu.co}}\hspace{3pt}\hspace{5pt}and\hspace{5pt}    
J. S. Alvarado\thanks{\url{jsalvaradog@unal.edu.co}}                                                            
\\
\small \textit{Departamento de Física, Universidad Nacional de Colombia, Carrera $30$ No. $45-03$, Bogotá D.C, Colombia}
}

\begin{document}
\maketitle
\begin{abstract}
We present an extension $U(1)_{X}$ to the Standard Model that reproduces the lepton mass structures determined by the experiments. In the charged sector, we introduced effective operators of dimension $n = 7$ to generate the mass of the electron, which is null at tree-level due to the $X$ charge. In the neutral sector, we added three sterile right-handed neutrinos and three Majorana neutrinos to generate the mass structure for the left-handed neutrinos, by the inverse seesaw mechanism. The model free parameters were fitted with the known mass eigenvalues $m_{e},\,m_{\mu},\,m_{\tau}$, and with the most recent results of a global analysis of the data from neutrino oscillation. From the adjustment of the free parameters, we obtained allowed regions for the Yukawa coupling set.
\end{abstract}


\section{Introduction}

Recently various experiments such as Super-Kamiokande \cite{wendell}, SAGE \cite{sage}, MINOS \cite{minos} and Double Chooz \cite{dchooz}, have confirmed neutrino oscillation using different neutrino sources (solar, atmospheric, of reactors and accelerators). Experimental data indicate that active neutrinos of the Standard Model (SM) have mass and their flavor fields are given by a combination of mass eigenstates (species)
\begin{equation}
\label{1-mezclaneutrinos}
    \ket{\nu_{L}^{a}}=\sum_{i=1,2,3}U_{ai}\ket{\nu_{L}^{i}},\quad\qquad a=e,\mu,\tau
\end{equation}
where $U\equiv U_{\text{PMNS}} = (V_{L}^{l})^{\dagger}V_{L}^{\nu}$ is the Pontecorvo-Maki-Nakagawa-Sakata (PMNS) matrix \cite{pontecorvo2, mns}. The matrices $V_{L}^{l}$ and $V_{L}^{\nu}$ diagonalize the mass matrix of charged leptons and active neutrinos, respectively. In general, if neutrinos are Dirac particles the PMNS matrix can be parameterized in terms of three mixing angles $\theta_{12},\,\theta_{13}\,\theta_{23}$ and one CP violating phase $\delta$ \cite{lipari}:
\begin{equation}
\label{1-PMNS}
    U_{\text{PMNS}}=\begin{pmatrix}
      c_{12}c_{13} & s_{12}c_{13} & s_{13}e^{-i\delta} \\
      -s_{12}c_{23}-c_{12}s_{23}s_{13}e^{i\delta} & c_{12}c_{23}-s_{12}s_{23}s_{13}e^{i\delta} & s_{23}c_{13} \\
      s_{12}s_{23}-c_{12}s_{23}s_{13}e^{i\delta} & -c_{12}c_{23}-s_{12}s_{23}s_{13}e^{i\delta} & c_{23}c_{13}
    \end{pmatrix},
\end{equation}
with $s_{ij}=\sin\theta_{ij}$ and $c_{ij}=\cos\theta_{ij}$. If neutrinos are Majorana particles, the PMNS matrix includes two additional phases ($\alpha,\,\beta$) but they have no influence on neutrino oscillation.

All data analyzed by the experiments mentioned and others can be described consistently by means of two non-equivalent arrangements for mass eigenvalues \cite{gonzalez}:
\begin{align}
\label{1-NO}
    &\text{Normal Ordering:}\qquad\quad \,\,m_{1}<m_{2}<m_{3}, \qquad\quad (\Delta m_{32}^{2}\simeq\Delta m_{31}^{2}>0),\\
\label{1-IO}
    &\text{Inverted Ordering:}\qquad\quad m_{3}<m_{1}<m_{2}, \qquad\quad (\Delta m_{32}^{2}\simeq\Delta m_{31}^{2}<0),
\end{align}
where the squared-mass differences are $\Delta m_{ij}^{2}=m_{i}^{2}-m_{j}^{2}$ $(i,j=1,2,3)$. From the latest global analysis reported by Esteban et al. \cite{esteban, nufit}, experimental values of squared-mass differences are listed in Table \ref{tab:1-nufit}. Esteban et al. also determined upper and lower limits for each component of the PMNS matrix
\begin{equation}
\label{1-expPMNS}
    |U|_{\text{PMNS}}^{SK,3\sigma}=\begin{pmatrix}
      0.801\rightarrow0.845 & 0.513\rightarrow0.579 &  0.143\rightarrow0.156 \\
      0.244\rightarrow0.499 & 0.505\rightarrow0.693 & 0.631\rightarrow0.768 \\
      0.272\rightarrow0.518 & 0.471\rightarrow0.669 & 0.623\rightarrow0.761 \\
    \end{pmatrix}.
\end{equation}
\begin{table}[h]
    \centering
    \begin{tabular}{c|c|c}
        & Normal Ordering (NO) & Inverted Ordering (IO) \\
        \hline \hline
        $\frac{\Delta m_{21}^{2}}{10^{-5}\,\text{eV}^{2}}$  &  $6.82\rightarrow8.04$ & $6.82\rightarrow8.04$ \\ 
        \hline
        $\frac{\Delta m_{3l}^{2}}{10^{-3}\,\text{eV}^{2}}$ & $2.430\rightarrow2.593$ &  $-2.574\rightarrow-2.410$  \\
    \end{tabular}
    \caption{Squared-mass differences at $3\,\sigma$ reported by Esteban et al., using Super Kamiokande data. $l=1$ for NO and $l=2$ for IO.}
\label{tab:1-nufit}
\end{table}

Seeing this, what theoretical frameworks can we use for the explanation of the neutrinos small mass and neutrino oscillation? The most studied method that answers our question is the seesaw mechanism, which adds right-handed neutrinos to the SM to give mass to left-handed neutrinos. Since the new energy scale associated with the new fields is high ($\sim 10^{14}-10^{16}$ GeV \cite{giunti}), the seesaw mechanism can not be tested by the experiments. Therefore in this model we use the inverse seesaw mechanism (ISS) which adds very light right-handed Majorana neutrinos to the SM, such that in the basis $(\nu_{L}, \nu_{R}^{C}, N_{R}^{C})$ the mass matrix has the form:
\begin{equation}
\label{1-fritzsch}
    \mathbb{M}_{\nu}=\begin{pmatrix} 0 & m_{\nu}^{T} & 0\\
                                     m_{\nu} & 0 & m_{N}^{T}\\
                                     0 & m_{N} & M_{N}
    \end{pmatrix},
\end{equation}
where the matrix block $m_{N}$ has component of the TeV scale order, $M_{N}$ is in KeV scale  and $m_{\nu}$ is in the electroweak scale. In this way the active neutrinos are obtained in sub-eV scale.

On the other hand, we can use the effective field theory as a theoretical framework to explain some experimental results. In this scenario we propose a dimensional expansion in the Lagrangian of the theory
\begin{equation}
\label{1-lagrangianoeff}
    \mathcal{L}=\mathcal{L}_{0}+\frac{\mathcal{L}_{1}}{\Lambda}+\frac{\mathcal{L}_{2}}{\Lambda^{2}}+\cdots,
\end{equation} 
where conventional renormalizable interactions are considered, $\mathcal{L}_{0}$, and non-renormalizable interactions are added, $\mathcal{L}_{n}$ ($n\geq1$) \cite{georgi}, which are described by operators of $n+4$ dimension suppressed by the new physics energy scale $\Lambda^{n}$. These effective operators add high-energy effects that can be measured on a low-energy scale.

In this work we use a next-to-minimal two Higgs double model (N2HDM) \cite{chen}, which adds elementary particles to the SM under a new symmetry $U(1)_{X}$. This symmetry is one of the most studied of the SM, as can be seen in the reference \cite{langacker}.

The paper is organized as follows. In the next section, we introduce the extension $U(1)_{X}$ to the SM, its particle content with their respective charge $X$ and hypercharge $Y$ values, which lead to zero the anomaly equations. In section $3$ we show how mass structures in the leptonic sector are predicted by the model. The mass of electron, which is massless at tree-level, is generated by effective operators of dimension $n = 7$ by introducing a Lambda scale, and the mass matrix of the active neutrinos is determined by the inverse seesaw mechanism. In section $4$, we present the parameter space of the model and the numerical formalism. The free parameters are fitted with the neutrino oscillation data available in NuFIT \cite{nufit}. Results are showed and analyzed in the section $5$.


\section{The $U(1)_{X}$ extension}

Under inclusion of a new non-universal gauge group $U(1)_{X}$, Alvarado et al. \cite{alvarado} and Mantilla et al. \cite{mantilla} proposed the next extension to the scalar sector of the SM:
\begin{table}[h]
    \centering
    \begin{tabular}{c|c|c|c}
       Scalar bosons & X & $\mathbb{Z}_{2}$ & Y \\
       \hline \hline 
       \multicolumn{4}{c}{Doublets} \\ 
       \hline \hline 
        $\phi_{1}=\begin{pmatrix} 
                \phi_{1}^{+} \\
                \frac{h_{1}+v_{1}+i\eta_{1}}{\sqrt{2}}
                \end{pmatrix}$ & +2/3 & + & +1 \\
        $\phi_{2}=\begin{pmatrix} 
                \phi_{2}^{+} \\
                \frac{h_{2}+v_{2}+i\eta_{2}}{\sqrt{2}}
                \end{pmatrix}$ & +1/3 & - & +1 \\
        \hline \hline
        \multicolumn{4}{c}{Singlet} \\
        \hline \hline 
        $\chi = \frac{\xi_{\chi}+v_{\chi}+i\zeta_{\chi}}{\sqrt{2}}$ & -1/3 & + & 0
    \end{tabular}
    \caption{Bosonic content of the model with their respective charge $X$, hypercharge $Y$ and parity $\mathbb{Z}_{2}$ values.}
    \label{tab:2-bosones}
\end{table}

The scalar doublets $\phi_{1},\,\phi_{2}$ have vacuum expectation values (VEV) that relate to the electroweak VEV by $v=\sqrt{v_{1}^{2}+v_{2}^{2}}$. The internal symmetry $\mathbb{Z}_{2}$ is introduced to obtain matrices with suitable textures. The scalar singlet $\chi$ with VEV $v_{\chi}$ is used for spontaneous symmetry breaking (SSB) of $U(1)_{X}$ and also for the mass generation of the exotic fermions in the model. We assume $v_{\chi}\gg v$ because $v_{\chi}$ gives mass to the gauge field $Z'_{\mu}$ associated to the symmetry $U(1)_{X}$, and from the non-observations of the LHC, there is a lower bound for the $Z'_{\mu}$ mass ($4.5\,\text{TeV}<M_{Z'}$).

The fermionic sector of the proposed model \cite{alvarado, mantilla} is presented in Table \ref{tab:2-fermiones}, where we use the following notation:
\begin{equation}
\label{2-notacionfermione}
    U^{1,2,3}=(u,c,t),\qquad D^{1,2,3}=(d,s,b),\qquad e^{e,\mu,\tau}=(e,\mu,\tau),\qquad \nu^{e,\mu,\tau}=(\nu^{e},\nu^{\mu},\nu^{\tau}).
\end{equation}
\begin{table}[h]
    \centering
    \begin{tabular}{c c c || c c c}
       Quarks & X & $\mathbb{Z}_{2}$ & Leptons & X & $\mathbb{Z}_{2}$\\
       \hline \hline
       $q_{L}^{1}=\begin{pmatrix} 
                    U^{1} \\
                    D^{1}
                    \end{pmatrix}_{L}$ & +1/3 & + & $l_{L}^{e}=\begin{pmatrix}
                    \nu^{e} \\
                    e^{e}
                    \end{pmatrix}_{L}$ & 0 & + \\
       $q_{L}^{2}=\begin{pmatrix} 
                    U^{2} \\
                    D^{2}
                    \end{pmatrix}_{L}$ & 0 & - & $l_{L}^{\mu}=\begin{pmatrix}
                    \nu^{\mu} \\
                    e^{\mu}
                    \end{pmatrix}_{L}$ & 0 & + \\
        $q_{L}^{3}=\begin{pmatrix} 
                    U^{3} \\
                    D^{3}
                    \end{pmatrix}_{L}$ & 0 & + & $l_{L}^{\tau}=\begin{pmatrix}
                    \nu^{\tau} \\
                    e^{\tau}
                    \end{pmatrix}_{L}$ & -1 & + \\
        \hline \hline
        $U_{R}^{1,3}$ & +2/3 & + & $e_{R}^{e,\tau}$ & -4/3 & - \\
        $U_{R}^{2}$ & +2/3 & - & $e_{R}^{\mu}$ & -1/3 & - \\
        $D_{R}^{1,2,3}$ & -1/3 & - & \\
        \hline \hline
        \multicolumn{6}{c}{Extension} \\
        \hline \hline
        $T_{L}$ & +1/3 & - & $\nu_{R}^{e,\mu,\tau}$ & +1/3 & - \\
        $T_{R}$ & +2/3 & - & $N_{R}^{e,\mu,\tau}$ & 0 & - \\
        $J_{L}^{1,2}$ & 0 & + & $E_{L}, \mathcal{E}_{R}$ & -1 & + \\
        $J_{R}^{1,2}$ & -1/3 & + & $E_{R}, \mathcal{E}_{L}$ & -2/3 & + \\
    \end{tabular}
    \caption{Fermionic content of the model with their respective charge $X$ and parity $\mathbb{Z}_{2}$ values.}
    \label{tab:2-fermiones}
\end{table}
\newpage
The charge $X$ of the symmetry $U(1)_{X}$ and hypercharge $Y$ assigned to the fermions are such that they set the anomaly equations to zero \cite{alvarado, mantilla, ordell}:
\begin{align}
\label{2-anomauno}
    [SU(3)_{C}]^{2}U(1)_{X}\rightarrow A_{C}&=\sum_{Q}[X_{Q_{L}}-X_{Q_{R}}],\\
\label{2-anomados}
    [SU(2)_{L}]^{2}U(1)_{X}\rightarrow A_{L}&=\sum_{l}X_{l_{L}}+3\sum_{Q}X_{Q_{L}},\\
\label{2-anomatres}
    [U(1)_{Y}]^{2}U(1)_{X}\rightarrow A_{Y^{2}}&=\sum_{l,Q}[Y_{l_{L}}^{2}X_{l_{L}}+3Y_{Q_{L}}^{2}X_{Q_{L}}]-\sum_{l,Q}[Y_{l_{R}}^{2}X_{l_{R}}+3Y_{Q_{R}}^{2}X_{Q_{R}}],\\
\label{2-anomacuatro}
    U(1)_{Y}[U(1)_{X}]^{2}\rightarrow A_{Y}&=\sum_{l,Q}[Y_{l_{L}}X_{l_{L}}^{2}+3Y_{Q_{L}}X_{Q_{L}}^{2}]-\sum_{l,Q}[Y_{l_{R}}X_{l_{R}}^{2}+3Y_{Q_{R}}X_{Q_{R}}^{2}],\\
\label{2-anomacinco}
    [U(1)_{X}]^{3}\rightarrow A_{X}&=\sum_{l,Q}[X_{l_{L}}^{3}+3X_{Q_{L}}^{3}]-\sum_{l,Q}[X_{l_{R}}^{3}+3X_{Q_{R}}^{3}],\\
\label{2-anomaseis}
    [\text{Grav}]^{2}U(1)_{X}\rightarrow A_{G}&=\sum_{l,Q}[X_{l_{L}}+3X_{Q_{L}}]-\sum_{l,Q}[X_{l_{R}}+3X_{Q_{R}}],
\end{align}
where $Q$ runs over quarks and $l$ runs over leptons with non-trivial values of $U(1)_{X}$. In order to obtain an anomaly-free model and use ordinary electric charges, we added new particles to the charged sector: two bottom quarks $J^{a}$, one top quark $T$ and two exotic leptons with electric charge $+1$. Inside the neutral sector, we introduced six sterile neutrinos (three of Dirac $\nu_{R}$ and three of Majorana $N_{R}$) to provide masses to the active neutrinos of the SM through the inverse seesaw mechanism.

At last, because we add the extra gauge boson $Z'_{\mu}$ to set $U(1)_{X}$ as a local symmetry, the covariant derivative of the model is:
\begin{equation}
\label{2-derivadacov}
    D_{\mu}=\partial_{\mu}+\frac{ig}{2}A_{\mu}^{\alpha}\sigma_{\alpha}+\frac{ig'}{2}YB_{\mu}+ig_{X}XZ'_{\mu},
\end{equation}
where $\sigma_{a}$ are the Pauli matrices ($a=1,2,3$). Moreover, the definition of electric charge given by Gell-Mann-Nishijima remains unchanged
\begin{equation}
\label{2-cargaelec}
    Q=\frac{1}{2}\left(\sigma_{3}+Y\right).
\end{equation}


\section{Leptonic sector}

\subsection{Charged leptons}

From the symmetry $U(1)_{X}\otimes\mathbb{Z}_{2}$, the most general Lagrangian for the charged leptons is given by \cite{alvarado}:
\begin{equation}
\label{3-cargadolag}
    \begin{split}
    -\mathcal{L}_{\text{Y,C}}&=\eta\anti{l^{e}_{L}}\Phi_{2}e_{R}^{\mu}+h\anti{l^{\mu}_{L}}\Phi_{2}e_{R}^{\mu}+\zeta\anti{l^{\tau}_{L}}\Phi_{2}e_{R}^{e}+H\anti{l^{\tau}_{L}}\Phi_{2}e_{R}^{\tau}\\
    &\qquad+q_{11}\anti{l^{e}_{L}}\Phi_{1}E_{R} +q_{21}\anti{l^{\mu}_{L}}\Phi_{1}E_{R}+g_{\chi_{E}}\anti{E_{L}}\chi E_{R}+g_{\chi_{\mathcal{E}}}\anti{\mathcal{E}_{L}}\chi^{*} \mathcal{E}_{R}+\text{h.c.}   
    \end{split}
\end{equation}
The lepton $\mathcal{E}$ is decoupled and gets mass $m_{\mathcal{E}}=g_{\chi_{\mathcal{E}}}v_{\chi}/\sqrt{2}$. When the symmetry breaks spontaneously, in the flavor basis $\va{\mathbf{E}}=(e^{e},e^{\mu},e^{\tau},E)^{T}$ we obtain the following $4\times4$ matrix mass \cite{alvarado}
\begin{equation}
\label{3-masainicial}
    M_{l}^{0}=
    \begin{pmatrix}
      0 & \frac{\eta v_{2}}{\sqrt{2}} & 0 & \frac{q_{11} v_{1}}{\sqrt{2}} \\
      0 & \frac{h v_{2}}{\sqrt{2}} & 0 & \frac{q_{21} v_{1}}{\sqrt{2}} \\
      \frac{\zeta v_{2}}{\sqrt{2}} & 0 & \frac{H v_{2}}{\sqrt{2}} & 0 \\
      0 & 0 & 0 & \frac{g_{\chi_{E}} v_{\chi}}{\sqrt{2}}
    \end{pmatrix}.
\end{equation}
The matrix \eqref{3-masainicial} has rank $r=3$, which implies that the lightest lepton (electron) is massless at tree-level. Therefore, we consider the following additional effective Lagrangian to the model \cite{alvarado}
\begin{equation}
\label{3-efflag}
    \begin{split}
    \mathcal{L}_{\text{effective}}&=\mathcal{O}_{ij}^{l}+\mathcal{O}_{\tau\mu}^{l}+\mathcal{O}_{Ej}^{l}+\mathcal{O}_{E\mu}^{l}+\mathcal{O}_{\tau E}^{l}\\
    &=\Omega_{ij}^{l}\carf{\chi^{*}}{\Lambda}^{3}\anti{l_{L}^{i}}\Phi_{2}e_{R}^{j}+\Omega_{\tau\mu}^{l}\carf{\chi}{\Lambda}^{3}\anti{l_{L}^{\tau}}\Phi_{2}e_{R}^{\mu}+\Omega_{Ej}^{l}\frac{\Phi_{2}^{\dagger}\Phi_{1}}{\Lambda}\anti{E_{L}}e_{R}^{j}\\
    &\qquad+\Omega_{E\mu}^{l}\frac{\Phi_{1}^{\dagger}\Phi_{2}\chi}{\Lambda^{2}}\anti{E_{L}}e_{\mu}^{j}+\Omega_{\tau E}^{l}\carf{\chi}{\Lambda}^{3}\anti{l_{L}^{\tau}}\Phi_{1}E_{R},
    \end{split}
\end{equation}
where $i=e,\mu$, $j=e,\tau$ and $\Lambda$ is the associated energy scale. Hence, the new mass matrix in the charged sector is
\begin{equation}
\label{3-masaeff}
    M_{l}=
    \begin{pmatrix}
    \Omega_{ee}^{l}\frac{v_{2}v_{\chi}^{3}}{4\Lambda^{3}} & \frac{\eta v_{2}}{\sqrt{2}} & \Omega_{e\tau}^{l}\frac{v_{2}v_{\chi}^{3}}{4\Lambda^{3}} & \frac{q_{11} v_{1}}{\sqrt{2}} \\
    \Omega_{\mu e}^{l}\frac{v_{2}v_{\chi}^{3}}{4\Lambda^{3}} & \frac{h v_{2}}{\sqrt{2}} & \Omega_{\mu\tau}^{l}\frac{v_{2}v_{\chi}^{3}}{4\Lambda^{3}} & \frac{q_{21} v_{1}}{\sqrt{2}} \\
    \frac{\zeta v_{2}}{\sqrt{2}} & \Omega_{\tau\mu}^{l}\frac{v_{2}v_{\chi}^{3}}{4\Lambda^{3}} & \frac{H v_{2}}{\sqrt{2}} & \Omega_{\tau E}^{l}\frac{v_{1}v_{\chi}^{3}}{4\Lambda^{3}} \\
    \Omega_{Ee}^{l}\frac{v_{1}v_{2}}{2\Lambda} & \Omega_{E\mu}^{l}\frac{v_{1}v_{2}v_{\chi}}{2\sqrt{2}\Lambda^{2}} & \Omega_{E\tau}^{l}\frac{v_{1}v_{2}}{2\Lambda} & \frac{g_{\chi_{E}} v_{x}}{\sqrt{2}}
    \end{pmatrix}.
\end{equation}
The effective operators $\mathcal{O}^{l}$ are of dimension $n=7$ and invariant under the symmetry of the model.

The rotation matrix $V_{L}^{l}$ that connects flavor states $\va{\mathbf{E}}$ with the mass eigenstates $\va{\mathbf{e}}=(e,\mu,\tau,E)^{T}$ through
\begin{equation*}
\label{3-rotizq}
    \va{E}_{L}=V_{L}^{l}\va{e}_{L},
\end{equation*}
can be rewritten as the product of two sub-rotations ($V_{L}^{l}\approx V_{L,1}^{l}V_{L,2}^{l}$) \cite{alvarado}
\begin{align}
\label{3-matrizrot1}
    V_{L,1}^{l}&=
    \begin{pmatrix}
    1 & 0 & 0 & \frac{q_{11} v_{1}}{\sqrt{2}m_{E}} \\
    0 & 1 & 0 & \frac{q_{21} v_{1}}{\sqrt{2}m_{E}} \\
    0 & 0 & 1 & r_{3} \\
    -\frac{q_{11} v_{1}}{\sqrt{2}m_{E}} & -\frac{q_{21} v_{1}}{\sqrt{2}m_{E}} & -r_{3} & 1
    \end{pmatrix},\\
\label{3-matrizrot2}
    V_{L,2}^{l}&=
    \begin{pmatrix}
    \ceu & \seu & r_{1} & 0 \\
    -\seu & \ceu & r_{2} & 0 \\
    -r_{1}\ceu+r_{2}\seu & -r_{2}\ceu-r_{1}\seu & 1 & 0 \\
    0 & 0 & 0 & 1
    \end{pmatrix},
\end{align}
with $\teu=\tan\theta_{e\mu}=\eta/h$ and $\tet=\tan\theta_{e\tau}=\zeta/H$. As a first approximation, we consider $m_{E}^{2}\approx g_{\chi_{E}}^{2}v_{\chi}^{2}/2\gg1$, in such a way that the effects of the exotic lepton $E$ are not considered at low energies (electroweak scale); thus $V_{L,1}^{l}$ can be approximated to identity. Then, we only consider the following matrix block $3\times3$ of $V_{L}^{l} \simeq V_{L,2}^{l}$:
\begin{equation}
\label{3-rotacioncargada}
    V_{L,3\times3}^{l}=
    \begin{pmatrix}
    \ceu & \seu & r_{1} \\
    -\seu & \ceu & r_{2} \\
    -r_{1}\ceu+r_{2}\seu & -r_{2}\ceu-r_{1}\seu & 1
    \end{pmatrix}.
\end{equation}
In this way, the mass eigenvalues are:
\begin{align}
\label{3-electron}
    m_{e}^{2}&\approx\frac{v_{2}}{4}\carf{v_{\chi}}{\Lambda}^{6}[\set(\Omega_{\mu\tau}^{l}\seu-\Omega_{e\tau}^{l}\ceu)+\cet(\Omega_{ee}^{l}\ceu-\Omega_{\mu e}^{l}\seu)]^{2},\\
\label{3-muon}
    m_{\mu}^{2}&\approx\frac{1}{2}(\eta^{2}+h^{2})v_{2}^{2},\\
\label{3-tauon}
    m_{\tau}^{2}&\approx\frac{1}{2}(\zeta^{2}+H^{2})v_{2}^{2},
\end{align}
and the parameters $r_{1},\,r_{2}$ have the form:
\begin{align}
\label{3-runo}
    r_{1}&=\frac{\set\Omega_{ee}^{l}+\cet\Omega_{e\tau}^{l}+\sqrt{\eta^{2}+h^{2}}\seu\Omega_{\tau\mu}^{l}}{2\sqrt{2}\sqrt{\zeta^{2}+H^{2}}}\carf{v_{\chi}}{\Lambda}^{3},\\
\label{3-rdos}
    r_{2}&=\frac{\set\Omega_{\mu e}^{l}+\cet\Omega_{\mu\tau}^{l}+\sqrt{\eta^{2}+h^{2}}\seu\Omega_{\tau\mu}^{l}}{2\sqrt{2}\sqrt{\zeta^{2}+H^{2}}}\carf{v_{\chi}}{\Lambda}^{3}.
\end{align}
The mass eigenvalues impose some restrictions on the initial parameter space of the model $\{\eta,h,\zeta,H,\Lambda\}$, likewise, these depend on the values selected for $\{v_{2},v_{\chi},\Omega^{l}\}$. According to the expressions \eqref{3-muon}-\eqref{3-tauon} the muon and tauon masses are in terms of $v_{2}$. Alvarado et al. also show that the mass of the boson $Z'$ are in terms of $v_{\chi}$ \cite{alvarado}. Consequently, we choose $v_{2}=2$ GeV and $v_{\chi}=7$ TeV to get the order of magnitude of the $\mu,\,\tau$ and $Z'$ masses . This selection also allows to establish the correct magnitude order of the left-handed neutrino masses, as shown in the next subsection. In turn, according to the equation \eqref{3-electron}, the mass of the electron provides the energy scale $\Lambda$ associated with the model.

\subsection{Neutral leptons}

The Yukawa couplings allowed for the neutral leptons are \cite{mantilla}:
\begin{equation}
\label{3-neutrolag}
    \begin{split}
           -\mathcal{L}_{\text{Y,N}}&=\yukawa{e}{e}\anti{l_{L}^{e}}\Tilde{\Phi}_{2}\nu_{R}^{e}+\yukawa{e}{\mu}\anti{l_{L}^{e}}\Tilde{\Phi}_{2}\nu_{R}^{\mu}+\yukawa{e}{\tau}\anti{l_{L}^{e}}\Tilde{\Phi}_{2}\nu_{R}^{\tau}+\yukawa{\mu}{e}\anti{l_{L}^{\mu}}\Tilde{\Phi}_{2}\nu_{R}^{e}+\yukawa{\mu}{\mu}\anti{l_{L}^{\mu}}\Tilde{\Phi}_{2}\nu_{R}^{\mu}+\yukawa{\mu}{\tau}\anti{l_{L}^{\mu}}\Tilde{\Phi}_{2}\nu_{R}^{\tau}\\
           &\qquad+h_{\chi i}^{\nu j}\anti{\nu_{R}^{iC}}\chi^{*}N_{R}^{j}+\frac{1}{2}\anti{N_{R}^{iC}}M_{N}^{ij}N_{R}^{j}+\text{h.c.},
      \end{split}
\end{equation}
where $\Tilde{\Phi}_{2}=i\sigma_{2}\Phi_{2}^{*}$ with $\sigma_{2}$ the second Pauli matrix.
After SSB and in the basis $\mathbf{N}_{L}=(\nu_{L}^{e,\mu,\tau},(\nu_{R}^{e,\mu,\tau})^{C},(N_{R}^{e,\mu,\tau})^{C})^{T}$ we obtain the following mass matrix:
\begin{equation}
\label{3-Nmasa}
        M_{\nu}=\begin{pmatrix}
                  0 & m_{D}^{T} & 0 \\
                  m_{D} & 0 & M_{D}^{T} \\
                  0 & M_{D} & M_{M}
               \end{pmatrix},
\end{equation}
with
\begin{equation}
        m_{D}^{T}=\frac{v_{2}}{\sqrt{2}}\begin{pmatrix}
                 \yukawa{e}{e} & \yukawa{e}{\mu} & \yukawa{e}{\tau}\\
                 \yukawa{\mu}{e} & \yukawa{\mu}{\mu} & \yukawa{\mu}{\tau}\\
                 0 & 0 & 0
               \end{pmatrix}
\end{equation}
the Dirac mass matrix between $\nu_{L}$ y $\nu_{R}$. $(M_{D})^{ij}=h_{\chi i}^{\nu j} v_{\chi}/\sqrt{2}$ is the Dirac matrix block between $\nu_{R}^{C}$ y $N_{R}$, and $(M_{M})^{ij}=(M_{N})^{ij}/2$ is the mass matrix of the Majorana neutrinos $N_{R}$ \cite{mantilla}. If we assume the hierarchy $M_{M}\ll m_{D}\ll M_{D}$, the mass of the active neutrinos is generated using the inverse see-saw mechanism \cite{mantilla, catano}. The $3\times3$ mass matrix of the active neutrinos is given by:
\begin{equation}
\label{4-mlight}
    m_{\text{light}}=m_{D}^{T}(M_{D})^{-1}M_{M}(M_{D}^{T})^{-1}m_{D},
\end{equation}
whereas the $6\times 6$ heavy term of the mechanism is
\begin{equation}
\label{4-mheavy}
    m_{\text{heavy}}\approx
    \begin{pmatrix}
    0 & M_{D}^{T}\\
    M_{D} & M_{M} 
    \end{pmatrix}.
\end{equation}

We consider for simplicity that $M_{D}$ is diagonal and $M_{M}$ is proportional to the identity:
\begin{equation}
    M_{D}=\frac{v_{\chi}}{\sqrt{2}}
    \begin{pmatrix}
    h_{N_{\chi_{e}}} & 0 & 0 \\
    0 & h_{N_{\chi_{\mu}}} & 0 \\
    0 & 0 & h_{N_{\chi_{\tau}}} \\
    \end{pmatrix},\qquad	M_{M}=\mu_{N}\mathbb{I}_{3\times3},
\end{equation}
then the mass matrix of the active neutrinos \eqref{4-mlight} is:
\begin{equation}
    m_{\text{light}}=\frac{\mu_{N}v_{2}^{2}}{h_{N_{\chi_{e}}}^{2}v_{\chi}^{2}}
    \begin{pmatrix}
    (\yukawa{e}{e})^{2}+(\yukawa{\mu}{e})^{2}\rho^{2} & \yukawa{e}{e}\yukawa{e}{\mu}+\yukawa{\mu}{e}\yukawa{\mu}{\mu}\rho^{2} & \yukawa{e}{e}\yukawa{e}{\tau}+\yukawa{\mu}{e}\yukawa{\mu}{\tau}\rho^{2} \\
    \yukawa{e}{e}\yukawa{e}{\mu}+\yukawa{\mu}{e}\yukawa{\mu}{\mu}\rho^{2} & (\yukawa{e}{\mu})^{2}+(\yukawa{\mu}{\mu})^{2}\rho^{2} & \yukawa{e}{\mu}\yukawa{e}{\tau}+\yukawa{\mu}{\mu}\yukawa{\mu}{\tau}\rho^{2} \\
    \yukawa{e}{e}\yukawa{e}{\tau}+\yukawa{\mu}{e}\yukawa{\mu}{\tau}\rho^{2} & \yukawa{e}{\mu}\yukawa{e}{\tau}+\yukawa{\mu}{\mu}\yukawa{\mu}{\tau}\rho^{2} & (\yukawa{e}{\tau})^{2}+(\yukawa{\mu}{\tau})^{2}\rho^{2}
    \end{pmatrix}
\end{equation}
where $\rho=h_{N_{\chi_{e}}}/h_{N_{\chi_{\mu}}}$. The matrix above has rank $r = 2$, which implies that an active neutrino is massless. In addition, we have an outer factor responsible for providing the mass scale of experiments for the left-handed neutrinos. As a first approximation, we consider that $\rho=1$ and $\yukawa{\mu}{e}=\yukawa{\mu}{\tau}=0$ without affecting the structure of $m_{\text{light}}$. Thus:
\begin{equation}
\label{4-mlightfinal}
    m_{\text{light}}=\frac{\mu_{N}v_{2}^{2}}{h_{N_{\chi_{e}}}^{2}v_{\chi}^{2}}
    \begin{pmatrix}
    (\yukawa{e}{e})^{2} & \yukawa{e}{e}\yukawa{e}{\mu} & \yukawa{e}{e}\yukawa{e}{\tau} \\
    \yukawa{e}{e}\yukawa{e}{\mu} & (\yukawa{e}{\mu})^{2}+(\yukawa{\mu}{\mu})^{2} & \yukawa{e}{\mu}\yukawa{e}{\tau} \\
    \yukawa{e}{e}\yukawa{e}{\tau} & \yukawa{e}{\mu}\yukawa{e}{\tau} & (\yukawa{e}{\tau})^{2}
    \end{pmatrix}.
\end{equation}
The matrix \eqref{4-mlightfinal} can be diagonalized by the expression
\begin{equation}
\label{3-rotmlight}
    V_{L}^{\nu T}m_{\text{light}}V_{L}^{\nu}=m_{\text{light}}^{\text{diag}}.
\end{equation}
The matrix $V_{L}^{\nu}$ contains the mixing angles that transform the flavor states $\nu_{L}^{e,\mu,\tau}$ into the mass eigenstates $\nu_{L}^{1,2,3}$.


\section{Parameter space}

By definition, the PMNS matrix is constructed from the rotation matrices of the charged \eqref{3-rotacioncargada} and neutral sector \eqref{3-rotmlight} \cite{pontecorvo2, mns} as follows
\begin{equation}
\label{4-PMNS}
     U_{\text{PMNS}}=(V_{L,3\times3}^{l})^{\dagger}V_{L}^{\nu}.
\end{equation}
The free parameters of the model are:
\begin{itemize}
    \item From the rotation matrix of the charged sector $V_{L,3\times3}^{l}$
\begin{equation}
\label{4-parametroscargados}
    \{\Lambda,\theta_{e\mu},\theta_{e\tau},\Omega_{ee}^{l},\Omega_{e\tau}^{l},\Omega_{\mu e}^{l},\Omega_{\mu\tau}^{l},\Omega_{\tau\mu}^{l}\},
\end{equation} 
    where the energy scale $\Lambda$ is fitted with the electron mass as we show below.
    
    \item From the rotation matrix of the neutral sector $V_{L}^{\nu}$
\begin{equation}
\label{4-parametrosneutros}
    \{\yukawa{e}{e},\yukawa{e}{\mu},\yukawa{e}{\tau},\yukawa{\mu}{\mu}\}.
\end{equation}
    \item For the outer factor in \eqref{4-mlightfinal} we choose $\mu_{N}v_{2}^{2}/h_{N_{\chi_{e}}}^{2}v_{\chi}^{2}=50$ meV and $h_{N_{\chi_{e}}}=0.1$ such that $\mu_{N}=12.25$ keV. This choice is made in this way to find the mass scale of the left-handed neutrinos, determined by the experiments.
\end{itemize}
The selection of angles $\theta_{e\mu},\,\theta_{e\tau}$ and the parameters $\Omega^{l}$ in \eqref{4-parametroscargados} define the following three case studies.


\subsection{Case 1: $\theta_{e\mu},\theta_{e\tau}\ll1;\, \Omega_{ee}^{l}\approx\Omega_{e\tau}^{l}\approx\Omega_{\mu\tau}^{l}\approx1$}

First, we can consider the following approximations $\seu,\set\ll1$ and $\ceu,\cet\approx1$, in such a way that the expressions \eqref{3-electron}, \eqref{3-runo} and \eqref{3-rdos} become:
\begin{align}
\label{4-rcasouno}
    r_{1}&=r_{2}=\frac{1}{2\sqrt{2}H}\carf{v_{\chi}}{\Lambda}^{3},\\
\label{4-ecasouno}
    m_{e}^{2}&\approx\frac{v_{2}^{2}}{4}\carf{v_{\chi}}{\Lambda}^{6},
\end{align}
where from the relation \eqref{3-tauon} and experimental value of the tauon mass $m_{\tau}\approx1.777$ GeV \cite{pdg}, we obtain $H\approx1.26$ GeV. On the other hand, from the expression \eqref{4-ecasouno} and the experimental value of the electron mass $m_{e}\approx0.000511$ GeV \cite{pdg}, we determine that the energy scale in this case is $\Lambda\approx87.56$ TeV. These results are valid for any value of $\Omega_{\mu e}^{l}$ and $\Omega_{\tau\mu}^{l}$.


\subsection{Case 2: $\theta_{e\mu},\theta_{e\tau}\sim\pi/4;\,\Omega_{\tau\mu}^{l}\approx1$}

From the selection $\seu,\set,\ceu,\cet\approx1/\sqrt{2}$, the relations \eqref{3-electron}, \eqref{3-runo} and \eqref{3-rdos} remain in the following ways considering two additional approximations:
\begin{itemize}
    \item \textbf{Type A} - $\Omega_{ee}^{l}\approx\Omega_{e\tau}^{l}\approx1$
\begin{align}
\label{4-runocasodosa}
        r_{1}&=\frac{1+h\seu}{2\sqrt{2}H}\carf{v_{\chi}}{\Lambda}^{3},\\
\label{4-rdoscasodosa}
        r_{2}&=\frac{\Omega_{\mu e}^{l}+\Omega_{\mu\tau}^{l}+2h\seu}{4\sqrt{2}H}\carf{v_{\chi}}{\Lambda}^{3},\\
\label{4-ecasodosa}
        m_{e}^{2}&\approx\frac{v_{2}^{2}}{8}\carf{v_{\chi}}{\Lambda}^{6}\seu^{2}(\Omega_{\mu\tau}^{l}-\Omega_{\mu e}^{l})^{2}.
\end{align}

Here the parameters $\Omega_{\mu e}^{l}$ and $\Omega_{\mu\tau}^{l}$ are free, but we choose to vary them within the interval $[0,10]$.

    \item \textbf{Type B} - $\Omega_{\mu e}^{l}\approx\Omega_{\mu\tau}^{l}\approx1$
\begin{align}
\label{4-runocasodosb}
        r_{1}&=\frac{\Omega_{ee}^{l}+\Omega_{e\tau}^{l}+2h\seu}{4\sqrt{2}H}\carf{v_{\chi}}{\Lambda}^{3},\\
\label{4-rdoscasodosb}
        r_{2}&=\frac{1+h\seu}{2\sqrt{2}H}\carf{v_{\chi}}{\Lambda}^{3},\\\
\label{4-ecasodosb}
         m_{e}^{2}&\approx\frac{v_{2}^{2}}{8}\carf{v_{\chi}}{\Lambda}^{6}\ceu^{2}(\Omega_{ee}^{l}-\Omega_{e\tau}^{l})^{2}.
\end{align}

For this selection the parameters $\Omega_{ee}^{l}$ and $\Omega_{e\tau}^{l}$ are free, and again we choose to vary them within the interval $[0,10]$.
\end{itemize}
The equations \eqref{3-muon}, \eqref{3-tauon} and experimental values of the muon and tauon masses, $m_{\mu}\approx0.1066$ GeV \cite{pdg} y $m_{\tau}\approx1.777$ GeV, leads to $h\approx0.05$ GeV and $H\approx0.89$ GeV.

For the energy scale $\Lambda$ determination, we used a Monte Carlo simulation with the equations \eqref{4-ecasodosa} (type A) and \eqref{4-ecasodosb} (type B). We vary the angle $\theta_{e\mu}$ in the interval $[0,\pi/2]$. Thus, the following histogram was obtained for the quantity $\Lambda/v_{\chi}$:
\begin{figure}[h]
    \centering
    \includegraphics[scale=0.4]{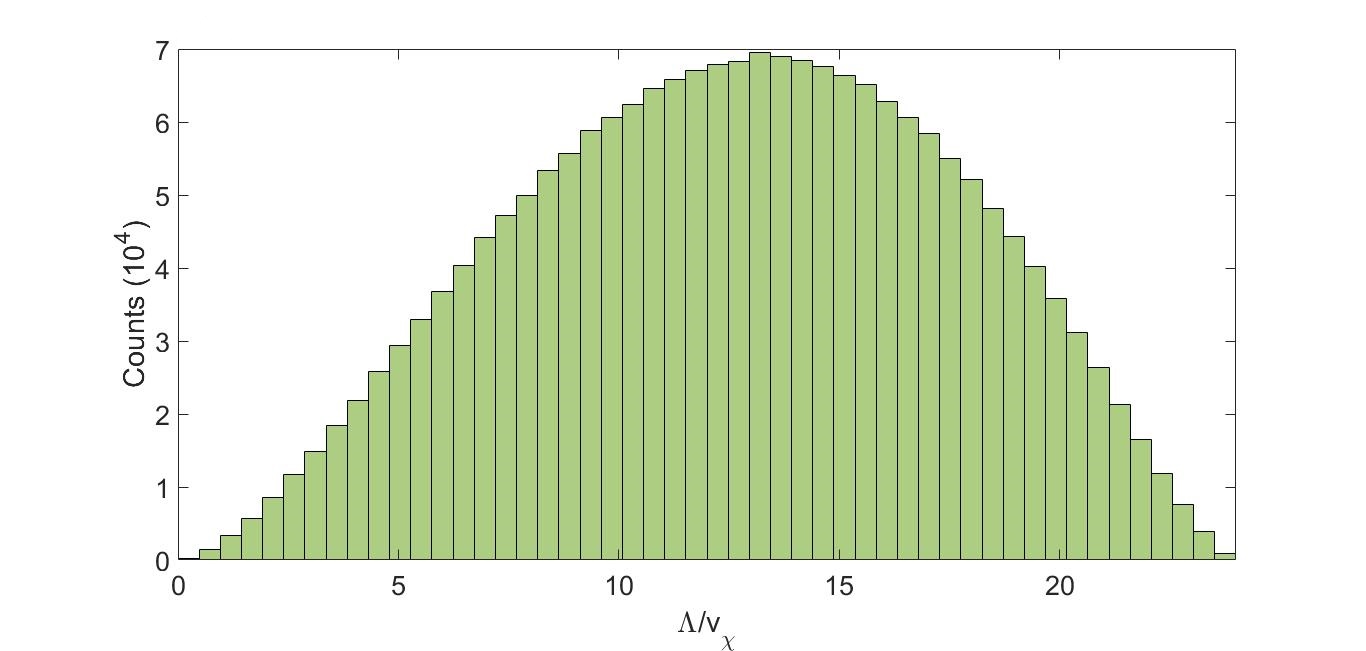}
    \caption{Accepted values distribution for the factor $\Lambda/v_{\chi}$ in the case $2$.}
\label{fig:4-scale}
\end{figure}
\newline
From the Figure \ref{fig:4-scale}, the mean value of $\Lambda/v_{\chi}$ is:
\begin{equation}
\label{4-escalaenergia}
    \expval{\frac{\Lambda}{v_{\chi}}}\approx12.72.
\end{equation}
This result is the same for the type A and type B selection. Then, the energy scale for both types is $\Lambda\approx89.05$ TeV.


\subsection{Case 3: $\theta_{e\mu}\ll1,\;\theta_{e\tau}\sim\pi/4$}

In this case, we choose $\seu\ll1$, $\ceu\approx1$ and $\set,\cet\approx1/\sqrt{2}$. Again we consider the two additional approximations made in the previous case. Then, the expressions \eqref{3-runo} and \eqref{3-rdos} take the form:
\begin{itemize}
    \item \textbf{Type C} - $\Omega_{ee}^{l}\approx\Omega_{e\tau}^{l}\approx1$
\begin{align}
\label{4-runocasotresa}
        r_{1}&=\frac{1}{2\sqrt{2}H}\carf{v_{\chi}}{\Lambda}^{3},\\
\label{4-rdoscasotresa}
        r_{2}&=\frac{\Omega_{\mu e}^{l}+\Omega_{\mu\tau}^{l}}{4\sqrt{2}H}\carf{v_{\chi}}{\Lambda}^{3}.
\end{align}

Here the parameters $\Omega_{\mu e}^{l}$ and $\Omega_{\mu\tau}^{l}$ are free, and we choose to vary them within the interval $[0,10]$.

    \item \textbf{Type D} - $\Omega_{\mu e}^{l}\approx\Omega_{\mu\tau}^{l}\approx1$
\begin{align}
\label{4-runocasotresb}
        r_{1}&=\frac{\Omega_{e e}^{l}+\Omega_{e\tau}^{l}}{4\sqrt{2}H}\carf{v_{\chi}}{\Lambda}^{3},\\
\label{4-rdoscasotresb}
         r_{2}&=\frac{1}{2\sqrt{2}H}\carf{v_{\chi}}{\Lambda}^{3}.
\end{align}

Again, we vary the free parameters $\Omega_{ee}^{l}$ and $\Omega_{e\tau}^{l}$ within the interval $[0,10]$.
\end{itemize}

For this case $H\approx0.89$ GeV. The final expressions of the electron squared mass are the same as those presented in the previous case: \eqref{4-ecasodosa} (type A) and \eqref{4-ecasodosb} (type B). Hence, in this case the energy scale is also $\Lambda\approx89.05$ TeV.\\


\section{Results and discussions}

We fitted the values of the Yukawa couplings $\yukawa{e}{e},\yukawa{e}{\mu},\yukawa{e}{\tau},\yukawa{\mu}{\mu}$, and the angle $\theta_{e\mu}$ using a Monte Carlo simulation. The accepted values were those that fit with the experimental ranges of the PMNS matrix components $|U|_{\text{PMNS}}^{\text{SK-atm},3\sigma}$, and the squared-mass differences for the Normal Ordering. The parameter space considered were $[0,1]$ for the Yukawa couplings $\yukawa{i}{j}$ and $[0,\pi/2]$ for $\theta_{e\mu}$.

The following values of the Yukawa parameters $\yukawa{i}{j}$ and angle $\theta_{e\mu}$ make the model compatible with the experimental neutrino oscillation results analyzed by Esteban et al. \cite{esteban}:

\begin{table}[h!]
    \centering
    \begin{tabular}{c||c c}
       $\theta_{e\mu}$ (rad) & $0.2842\rightarrow0.4182$ & $0.4931\rightarrow0.5549$ \\
       \hline \hline
       $\yukawa{e}{e}$  & $0.3604\rightarrow0.4353$ & $0.2054\rightarrow0.2706$ \\
       $\yukawa{e}{\mu}$  & $0.3972\rightarrow0.5676$ & $0.4651\rightarrow0.5982$ \\
       $\yukawa{e}{\tau}$  & $0.6425\rightarrow0.8123$ & $0.6272\rightarrow0.8929$ \\
       $\yukawa{\mu}{\mu}$  & $0.4665\rightarrow0.5442$ & $0.4992\rightarrow0.5503$ \\
     \end{tabular}
    \caption{Domain of the free parameters $\yukawa{i}{j}$, $\theta_{e\mu}$ that fits with the experimental neutrino oscillation data (SK-atm $3\sigma$) reports by Esteban et al. \cite{esteban}, for NO scheme.}
    \label{tab:6-NO}
\end{table}

As we mentioned earlier, our analysis only contemplates the Normal Ordering. Regardless of the case study, the values of the Table \ref{tab:6-NO} define the same compatibility region between model and experimental data.

\begin{figure}[h!]
        \centering
        \begin{subfigure}[b]{0.475\textwidth}
            \centering
            \includegraphics[width=\textwidth]{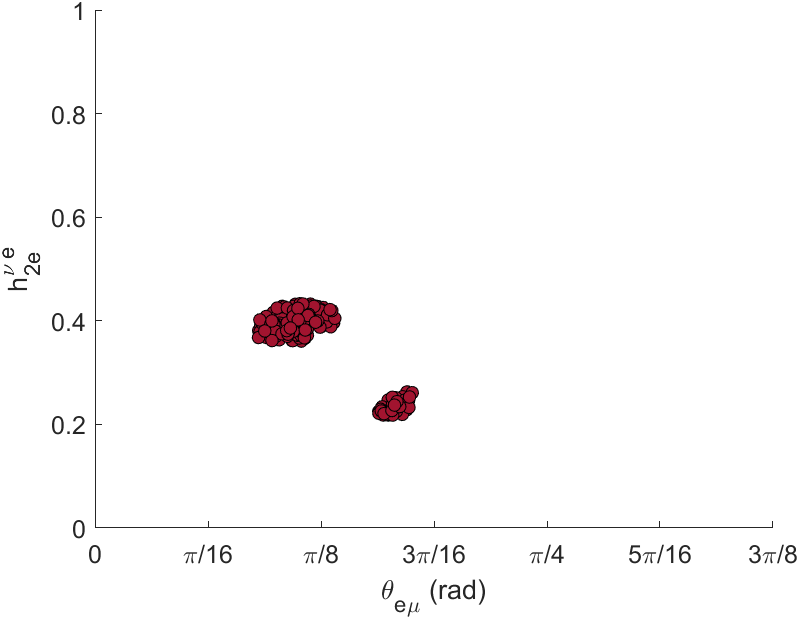}
            \caption[]%
            {{\small $\yukawa{e}{e}$}}    
            \label{fig:NOee}
        \end{subfigure}
        \hfill
        \begin{subfigure}[b]{0.475\textwidth}  
            \centering 
            \includegraphics[width=\textwidth]{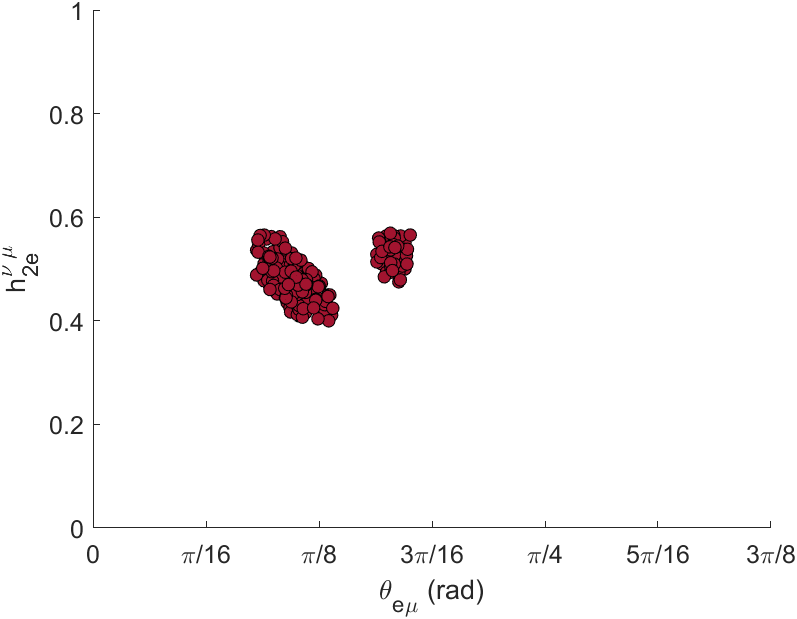}
            \caption[]%
            {{\small $\yukawa{e}{\mu}$}}    
            \label{fig:NOeu}
        \end{subfigure}
        \vskip\baselineskip
        \begin{subfigure}[b]{0.475\textwidth}   
            \centering 
            \includegraphics[width=\textwidth]{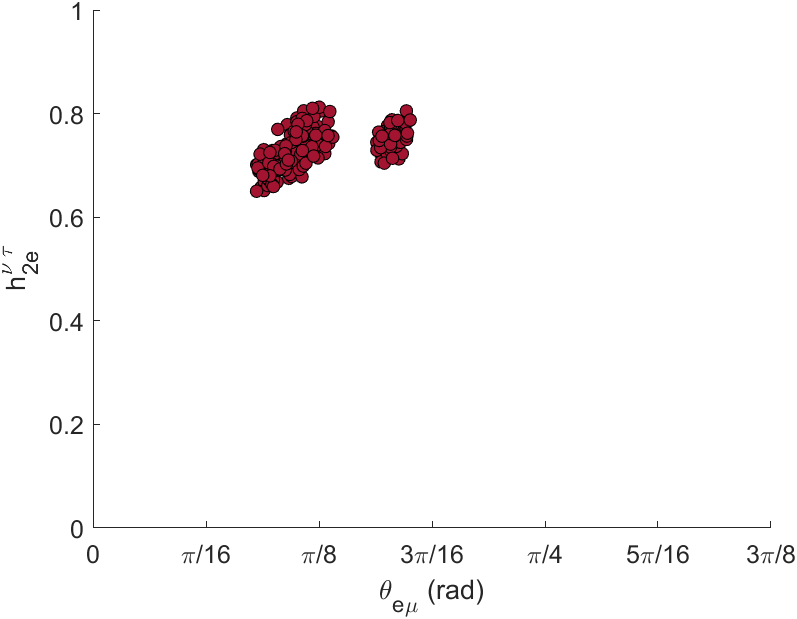}
            \caption[]%
            {{\small $\yukawa{e}{\tau}$}}    
            \label{fig:NOet}
        \end{subfigure}
        \hfill
        \begin{subfigure}[b]{0.475\textwidth}   
            \centering 
            \includegraphics[width=\textwidth]{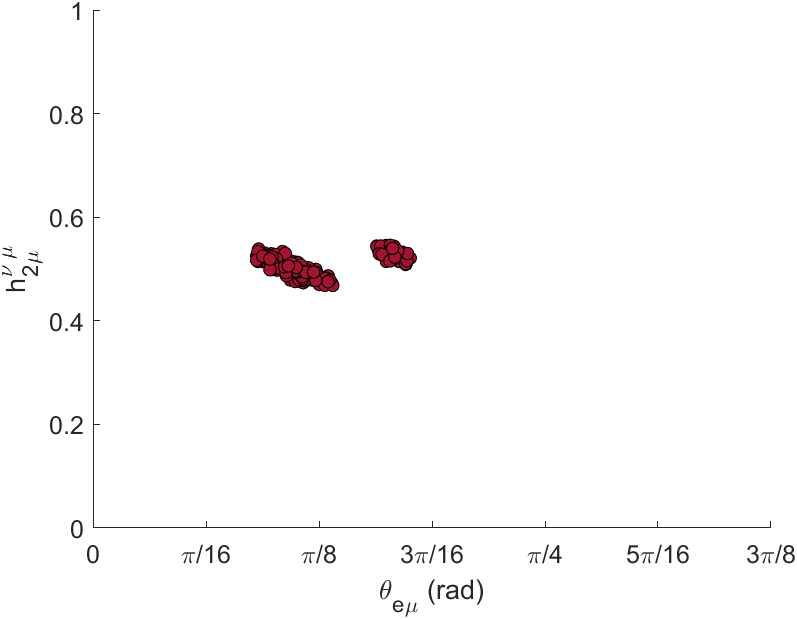}
            \caption[]%
            {{\small $\yukawa{\mu}{\mu}$}}    
            \label{fig:NOuu}
        \end{subfigure}
        \caption[]
        {\small Compatibility regions between model and experimental data available in NuFIT \cite{nufit}, for NO scheme.} 
        \label{fig:5-NO}
\end{figure}
\newpage
As we discuss in Sec. $3.2$ the model contains one massless active neutrino, then for NO scheme we selected $m_{1}=0$. The other masses $m_{2}$ and $m_{3}$ are given by their mean values, calculated from each set of parameters $\yukawa{i}{j}$ allowed by the Monte Carlo simulation. We compare the squared-mass differences calculated in the model with the experimental ranges given by Esteban et al. \cite{esteban}, for NO scheme:

\begin{table}[h]
    \centering
    \begin{tabular}{c|c|c}
        & NuFIT $3\sigma$ & $U(1)_{X}$ Model \\
        \hline \hline
        \multicolumn{3}{c}{Normal ordering}\\
        \hline \hline
        $\frac{\Delta m_{21}^{2}}{10^{-5}\,\text{eV}^{2}}$  &  $6.82\rightarrow8.04$ & $7.42$ \\
        $\frac{\Delta m_{31}^{2}}{10^{-3}\,\text{eV}^{2}}$  &  $2.430\rightarrow2.593$ & $2.500$ \\
    \end{tabular}
    \caption{Squared-mass differences in the model compared with the experimental values (SK-atm $3\sigma$) available in NuFIT \cite{nufit}, for NO scheme.}
    \label{tab:5-masses}
\end{table}

\newpage
\section{Conclusions}

We present a free-anomaly model that reproduces the lepton mass structures with few free parameters. The $\mu$ and $\tau$ leptons have mass at the electroweak scale and the exotics leptons $E,\,\mathcal{E}$ acquire mass at the $v_{\chi}$ scale. The electron obtains mass from effective operators of dimension $n = 7$, which requires introducing a new energy scale $\lambda$, and new Yukawa parameters, denoted by $\Omega_{ij}^{l}$. The equation \eqref{3-electron} is a complete expression for the mass of the electron in terms of the model free parameters. In the neutral sector, we generate the mass structure for the left-handed neutrinos from the addition of three sterile right-handed neutrinos, three Majorana neutrinos and using the inverse seesaw mechanism. We define three different cases of interest, varying the parameters $\{\theta_{e\mu},\,\theta_{e\tau},\,\Omega_{ij}^{l}\}$, in order to examine different regions of model free parameters and simplify expressions \eqref{3-electron}, \eqref{3-runo}, \eqref{3-rdos}. For these three cases we obtain an average value for the new energy scale $\Lambda\sim89\,\text{TeV}$, where we chose $v_{\chi}\simeq7\,\text{TeV}$ taking into account the lower bound for the $Z'_{\mu}$ mass from the LHC non-observations. In any case of interest, we establish the same compatibility regions between the model free parameters $(\yukawa{e}{e}, \yukawa{e}{\mu}, \yukawa{e}{\tau}, \yukawa{\mu}{\mu}, \theta_{e\mu})$ and the most recent experimental data from neutrino oscillation, using the squared-mass differences $\Delta m^{2}$ and the mixing angles in the PMNS matrix. Hence, we determine the free parameters values that recreate the experimental mass structure in the neutral sector, for Normal Ordering.
We do not find compatibility regions or the parameters in the case of Inverted Ordering.


\bibliographystyle{gerunsrt}
\bibliography{ref}

\end{document}